%% file: paper.tex
\documentclass[conference, nofonttune, english, letter]{ieeeconf}
\usepackage[protrusion=true,expansion=true]{microtype} 
\IEEEoverridecommandlockouts                              

\usepackage[pdftex]{graphicx}
\DeclareGraphicsExtensions{.pdf}
\graphicspath{{./media/}}
\usepackage{color}

\usepackage[cmex10]{amsmath}
\usepackage{amsthm}
\usepackage{amssymb}
\usepackage{algorithm, algorithmic}
\usepackage{balance}

\usepackage{array}
\usepackage{mdwmath}
\usepackage{mdwtab}
\usepackage{eqparbox}

\usepackage[british]{babel}

\usepackage[caption=false,font=footnotesize]{subfig}

\usepackage{dblfloatfix}

\usepackage{cite}

\usepackage{url}

\theoremstyle{definition}

\usepackage{amssymb, amsmath, latexsym, amsfonts, mathrsfs, amsbsy}
\usepackage{color}
\usepackage{bm}

\def \triset {\mathcal{T}}
\def \triindex {\tau}

\input{./macros.tex}

\title{\LARGE \bf
Experimental Validation of Stable Coordination for\\Multi-Robot Systems with Limited Fields of View using a Portable Multi-Robot Testbed\\{\large\normalfont EXTENDED ABSTRACT}
}

\author{Pratik Mukherjee$^{1}$, Matteo Santilli$^{2}$, Andrea Gasparri$^{2}$ and Ryan K. Williams$^{1}$%
\thanks{$^{1}$P. Mukherjee and R.K. Williams are with Electrical and Computer Engineering Department,
        Virginia Polytechnic Institute and State University, Blacksburg, VA USA,
        {\tt\small \{mukhe027, rywilli1\}@vt.edu}}%
\thanks{$^{2}$M. Santilli and A. Gasparri are with the Engineering Department, Roma Tre University
        Roma, 00146, Italy,
        {\tt\small matteo.santilli@uniroma3.it, gasparri@dia.uniroma3.it}}%
}

\begin{document}

\maketitle
\thispagestyle{empty}
\pagestyle{empty}


\section{Introduction}
In this paper, we address the problem of stable coordinated motion in multi-robot systems with limited fields of view (FOVs). These problems arise naturally for multi-robot systems that interact based on sensing, such as our case study of multiple unmanned aerial vehicles (UAVs) each equipped with several cameras that are used for detecting neighboring UAVs. 
 Related work can be separated into two topics:  asymmetric motion control and applications of perception in multi-robot systems. 
 Topology control for directed graphs is demonstrated in recent examples \cite{Asadi2016-wj,Sabattini2015-ln,Gasparri:CDC:2017:1} focusing on overcoming the theoretical shortcuts that are lost when the symmetry assumption is broken. Several recent works have exploited advanced perception in multi-robot systems, with examples including \cite{tokekar2014multi} which achieves multi-target tracking with camera-equipped unmanned aerial vehicles (UAVs),  \cite{Montijano2016-jv} which applies collaborative structure from motion for UAV formation control and \cite{kumar16-2} which demonstrates a distributed optimization framework for multi-robot collaborative tasks using vision. 

In this regard, our contribution is twofold.  First, we extend our framework \cite{Gasparri:CDC:2017:1} to study stable motion and distributed topology control for multi-robot systems with limited FOVs. Then, we provide experimental results with a team of DJI Matrice 100 UAVs performing motion control with limited FOVs to demonstrate the proposed control framework.

\section{Stable Directed Coordination with FOVs} \label{sec:sdcf}

Potential-based control design is a commonly used framework for controlling multi-robot systems~\cite{KODITSCHEK:1990,Zavlanos:2007,Ji:TRO:2007,Dimarogonas:2008,Zavlanos:2009,Tanner:TAC:2007, Williams:2013bh, Williams:2015:ICRA, Gasparri:TRO:2017:1}.  The basic idea is to encode the energy of a system as a potential function \mbox{$V(\mathbf{x}(t)) \in \mbb{R}_{+}$} such that the desired configurations of the multi-robot system correspond to critical points. For a review of stable motion framework for multi-robot systems with directed interactions we refer the reader to \cite{Gasparri:CDC:2017:1}. Notably, the underlying assumption of this stable motion with limited FOV framework is that each robot has its own proximity-limited  communication and sensing capability described by two  radii $\rho_{i,c}, \rho_{i,s} \in \mbb{R}_+, \forall \, i \in \{1, \ldots, n\}$,  within which sensing and communication can occur for each robot, respectively. 


In this regard, let us denote with $\mathcal{G}^{\text{FOV}}_s = \{ \mathcal{V}, \mathcal{E}^{\text{FOV}}_s\}$ the interaction graph encoding  pairwise  sensing interactions with limited field of view with node set $ \mathcal{V}\triangleq [v_1,...,v_n] $ and edge set $ \mathcal{E}^{\text{FOV}}_s\subseteq \mathcal{V} \times \mathcal{V} $.   At this point, let us introduce an  extended state~$\mathbf{x_i}$ defined as 
%
\begin{equation*}
\begin{aligned}
\mathbf{x_i} &= \left [ {x_i^\circ}^T, {x_i^{\triangleleft,1}}^T, {x_i^{\triangledown,1}}^T, {x_i^{\triangleright,1}}^T , \right . \ldots \\
& \hspace{3cm} \ldots, \, \left. {x_i^{\triangleleft,m_i}}^T, {x_i^{\triangledown,m_i}}^T, {x_i^{\triangleright,m_i}}^T \right ]^T
\end{aligned}
\end{equation*}
for each robot composed of the robot location itself, that is $x_i^\circ=x_i$, and a set of virtual points $\{ x_i^{\triangleleft,k}, x_i^{\triangledown,k}, x_i^{\triangleright,k}\}$ that move as if they were rigidly attached to a robot~$i$ for each FOV.  Note that, the position of each set of virtual points $\{x_i^{\triangleleft,k}, \, x_i^{\triangledown,k}, \, x_i^{\triangleright,k} \}$ is defined according to the orientation  $\theta_{i}^{k}$ of the FOV to which such set is associated,  that is  
$x_i^{\triindex,k} = R_i^{\triindex,k}(x_i) \, x_i + t_i^{\triindex,k}(x_i)$ where $\{ R_i^{\triindex,k}(x_i), t_i^{\triindex,k}(x_i) \}$ are pairs of proper rotation matrices and translation vectors and $\triindex$ is an element of the set $\triset \triangleq \{ \triangleleft, \triangledown, \triangleright \}$ denoting the virtual points. 

\begin{figure}
\centering
    \includegraphics[width=0.8\columnwidth]{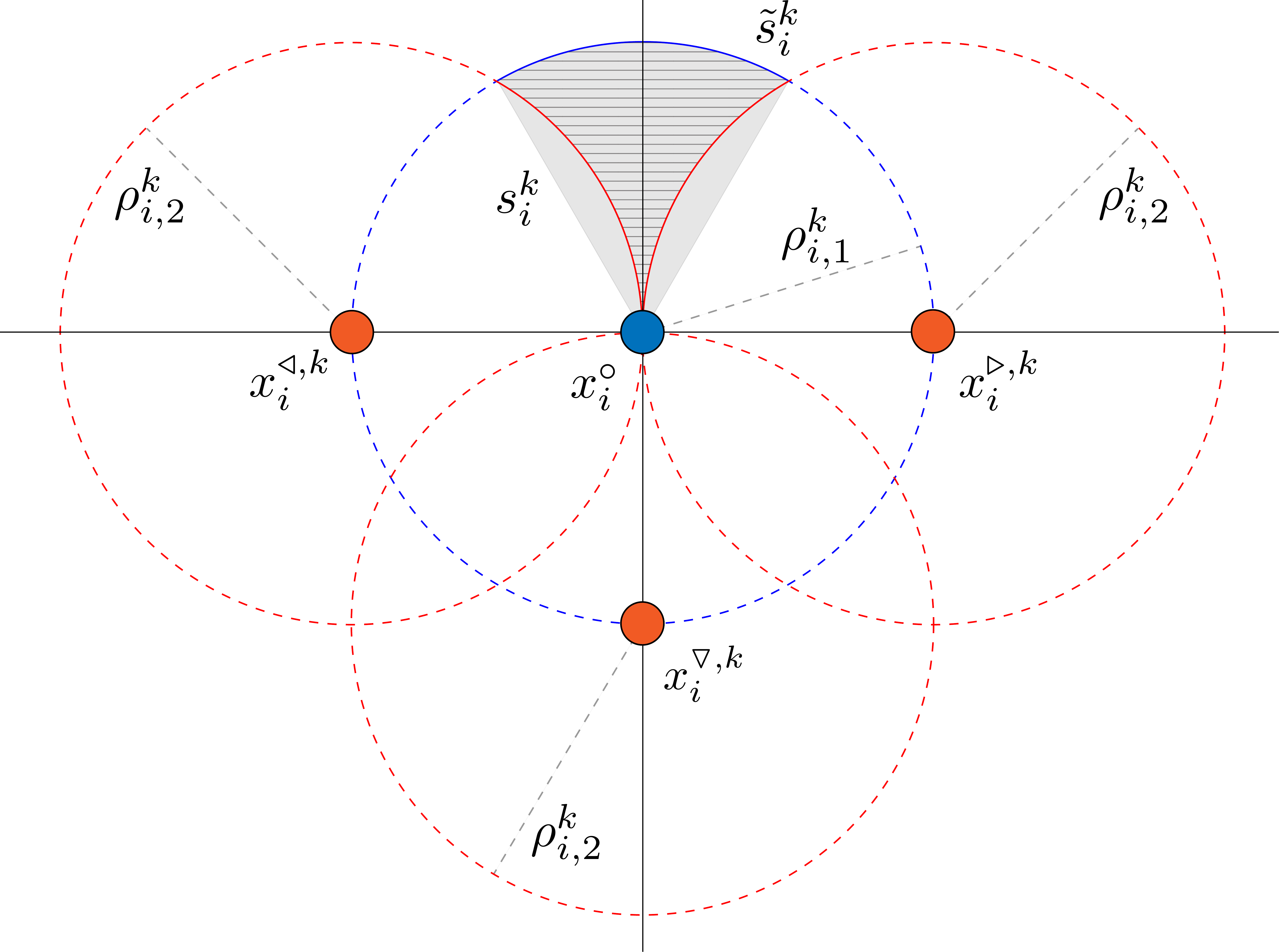}
    \caption{Approximation $\tilde{s}_i^k$ of the $k$-th circular sector $s_i^k$  for the limited field of view of an robot~$i$ by means of the set of logical conditions  given in~\eqref{eq:fov:1}.}
    \label{fig:approx_sensing}
\end{figure}

At this point, for each robot $i$ we can introduce an approximation $\tilde{s}_i^k$ of the $k$-th FOV (circular or spherical sector) $s_i^k$ as 
%
%
%
\begin{equation} \label{eq:fov:1}
 \tilde{s}_i^k = \left \{ x \in \mathbb{R}^d: 
\| x_i^\circ - x \| \leq \rho_{i,1}^k  \; \land \;
\| x_i^{\triindex,k} - x \| \geq \rho_{i,2}^k   \\
\right \}
\end{equation}
for all $\triindex \in \triset$ and $\rho_{i,1}^k, \, \rho_{i,2}^k  \in \mathbb{R}_+$ two radii chosen in such a way to approximate the $k$-th FOV(circular or spherical sector) of the $i$-th robot as defined by $s(x_i, \theta_i^k, \alpha_i^k, \rho_{i,s}^k)$, where $\land$ is the logical ``and'' operator. Therefore, it follows that 
given two robots~$i$ and~$j$ with state~$x_i$ and~$x_j$ respectively, we say that the robot~$j$  is within the  limited sensing field of view of robot~$i$ if there exists at least one approximation\footnote{If there is more than one circular (spherical) sector for which $x_j \in \tilde{s}_i^k$ then we assume robot $i$ locally selects the best one according to some sensing metric.  This guarantees that our sensing graph does not become a multigraph.} $\tilde{s}_i^k$ with $k \in \{1, \ldots, \,m_i\}$, such that $x_j \in \tilde{s}_i^k$. The reader is referred to Figure~\ref{fig:approx_sensing} for a graphical representation of the set of logical  conditions given in~\eqref{eq:fov:1}.

To   approximate any desired pairwise  sensing interaction with limited field of view,  the idea is that for each edge $(i,j) \in \mathcal{G}^{\text{FOV}}_s$ we can use a set of virtual points $ \{x_i^{\triangleleft,k}, \, x_i^{\triangledown,k}, \, x_i^{\triangleright,k}  \}$ along with the actual robot location $x_i^\circ$ to describe the desired interaction by means of a proper combination of  gradients. This allows us to derive a modeling of the multi-robot system with limited field of view, which we will refer to as the \emph{extended system} that is amenable to the theoretical framework in \cite{Gasparri:CDC:2017:1}.

Let us now consider the maintenance of a desired topological property  $\mathbb{P}$ as the design objective for the pairwise directed sensing interaction with limited field. More specifically, to the scope of this paper, let us assume the topological property of interest to be the maintenance of a directed link $(i,j) \in \mathcal{E}^{\text{FOV}}_s$. Notably, this objective can be translated in a setting with limited field of view by considering the following extended dynamics of each robot $i$:
\begin{equation}\label{eq:ext_dyn:1}
\begin{aligned}
\dot{x}_i^\circ &=  \underbrace{- \sum_{j \in \mathcal{N}_i^+} \left ( \nabla_{x_i^\circ} V_{{ij}}^{\circ}  + \sum_{\triindex \in \triset} \nabla_{x_i^{\triindex,k_j} } V_{{ij}}^{\triindex,k_j}  \right ) }_{u_i} \\
\end{aligned}
\end{equation}
with $\dot{x}_i^{\triindex,q} = u_i, q = 1,\ldots,m_i$ for each virtual point $\triindex \in \triset$, where $\mathcal{N}_i^+ = \left  \{ j \in \mathcal{V}, \, | \, (i,j) \in \mathcal{E}^{\text{FOV}}_s \right \} $ is defined according to~\eqref{eq:fov:1}, $k_j$ denotes the index $k$ for which $x_j \in \tilde{s}_i^k$ with $k \in 1, \, \ldots, m_i$ and the potentials $V_{{ij}}^{\circ}(\| x_i^\circ - x_j \|)$, $V_{{ij}}^{\triindex,k_j}(\| x_i^{\triindex}  - x_j \|)$ can be chosen such that
\begin{equation}
\begin{aligned}
& V_{{ij}}^{\circ,k_j}(\| x_i^\circ - x_j \|) \rightarrow \infty  \quad \text{as } \| x_i^\circ - x_j \| \rightarrow \rho_{i,1}^k,\\
& V_{{ij}}^{\triindex,k_j}(\| x_i^\triindex - x_j \|) \rightarrow \infty  \quad \text{as } \| x_i^\triindex - x_j \| \rightarrow \rho_{i,2}^k.\\
\end{aligned}
 \end{equation}
Interestingly,  two things can be noticed from~\eqref{eq:ext_dyn:1}:  i) the actual dynamics of the robot $x_i^\circ$ is influenced by the interactions of its $m_i$ sets of virtual points $\{ {x}_i^{\triangleleft,k},  {x}_i^{\triangledown,k}, {x}_i^{\triangleright,k}\}$, and ii) the dynamics of the $m_i$ sets of virtual points $\{\dot{x}_i^{\triangleleft,k}, \dot{x}_i^{\triangledown,k}, \dot{x}_i^{\triangleright,k}\}$  are identical to the actual dynamics of the robot  $\dot{x}_i^\circ$ being them rigidly attached to it. 

We can now study the stability of a multi-robot system $\mathbf{x} = [x_1^T, \, \ldots, \, x_n^T]^T$ with limited field of view,  by checking the stability of its extended version $\overline{\mathbf{x}} = [ \mathbf{x_1}^T, \, \ldots, \, \mathbf{x_n}^T]^T$.  To this end,  starting from the interaction graph  $\mathcal{G}^{\text{FOV}}_s = \{ \mathcal{V}, \mathcal{E}^{\text{FOV}}_s\}$, which encodes  the pairwise interactions with limited field of view, we require a systematic way for constructing  the interaction graph $\overline{\mathcal{G}}^{\text{FOV}}_s = \{ \overline{\mathcal{V}}^{\text{FOV}}, \, \overline{\mathcal{E}}^{\text{FOV}}_s   \}$ that encodes the equivalent pairwise interactions with limited field of view for its modeling based on the extended system. Indeed, this will permit to build the  incidence matrix  $\overline{\mathcal{B}} $ and the directed incidence matrix $\overline{\mathcal{B}}_+ $ associated to the graph $\overline{\mathcal{G}}^{\text{FOV}}_s$ which are required to check the stability of the system  by inspecting its extended system as per Theorem $3.1$ in \cite{Gasparri:CDC:2017:1}.

Intuitively, the idea is to: i) consider an extended state with $4 | \mathcal{E}_s|$ virtual points taken as replica of the actual agents; ii) perform a suitable algebraic manipulation of the extended state to zero out portions of the contributions corresponding to non-interacting virtual points  and iii) finally apply Theorem $3.1$ from \cite{Gasparri:CDC:2017:1} on the resulting extended states, yielding guaranteed stability. 

\section{Portable Multi-Robot Experimental Setup } \label{sec:experiments}

\begin{figure}[t!]
\centering
    \includegraphics[width=0.47\textwidth]{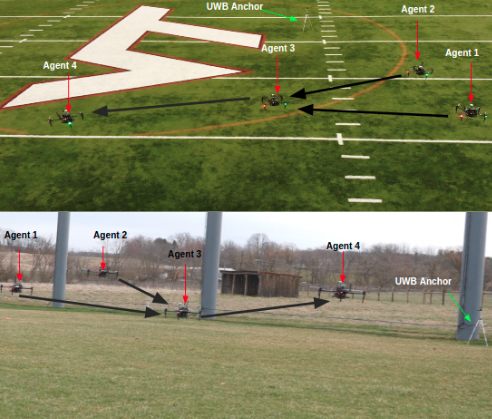}
    \caption{Four DJI Matrice 100s forming and maintaining a preselected stable directed interaction graph in  outdoor (bottom) and indoor(top) environments.}
    \label{fig:M100}
\end{figure}

\begin{figure}[t!]
\centering
\includegraphics[width=0.47\textwidth]{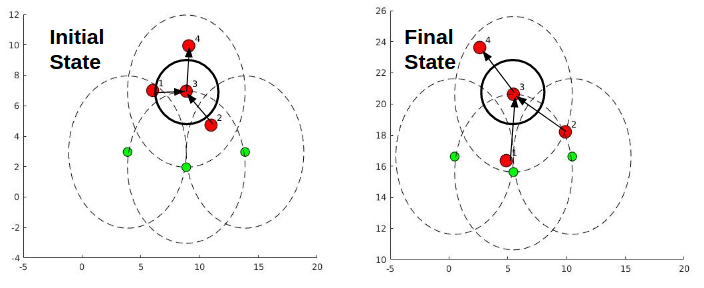}
    \caption{Initial and final states of the all agents(red circles) in an outdoor experiment with agent 4 in agent 3's FOV(one forward facing circular sector) represented by FOV radii of the three virtual points(green circles) of agent 3 and the collision radius represented by solid black circle around agent 3.}
    \label{fig:state}
\end{figure}
We deployed a team of DJI Matrice 100 UAVs and used an ultra-wideband (UWB) system, Pozyx~\cite{8373379}, for localization of the UAVs to control a stable FOV topology according to our theoretical results.
  To conduct experiments of topology control using onboard UWB localization, we place six Pozyx anchor UWB nodes in the environment.  Individual UWB tags are then mounted on each of the UAVs from which the position measurement is obtained. In the experiments conducted, we demonstrate limited FOV topology control of four UAVs operating in an area of $30$m $\times$ $20$m. A stable directed interaction graph was preselected as represented in Figures \ref{fig:M100}and \ref{fig:state}. This graph is maintained by all agents during experimentation.Here, we present the results from an outdoor experiment conducted at the Virginia Tech Drone Park facility.
 \\

\subsubsection{Outdoor Drone Park }
    \begin{figure}[t!]
\centering
\includegraphics[width=0.47\textwidth]{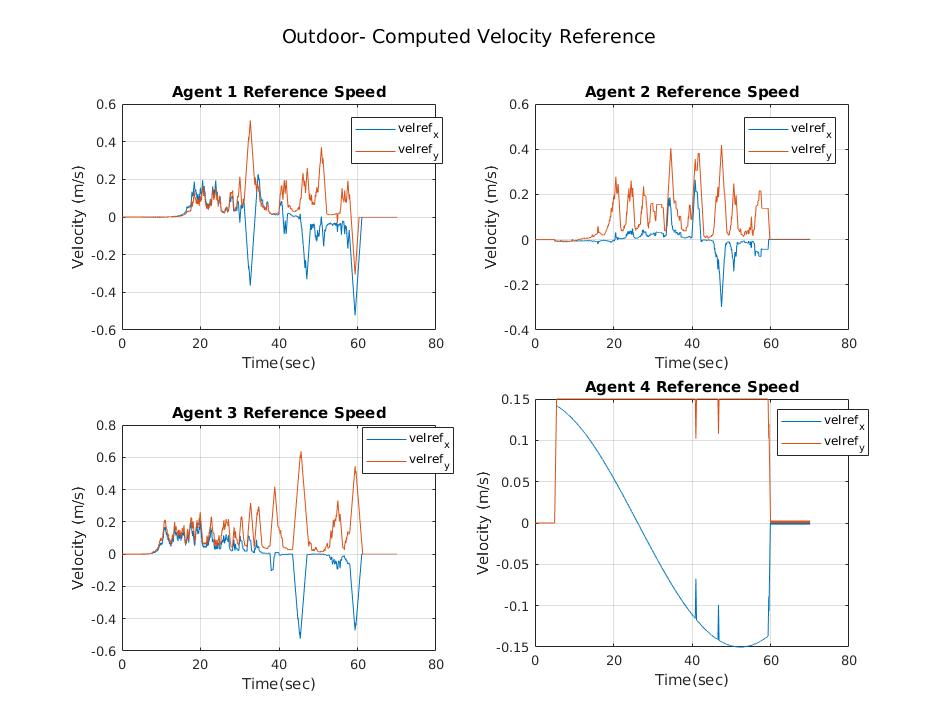}
    \caption{Computed velocity reference from FOV controller for agents 1,2 and 3 from outdoor experiment with agent 4 receiving a predetermined velocity.}
    \label{fig:out_vel_ref}
\end{figure}
\begin{figure}
\centering
    \includegraphics[width=0.47\textwidth]{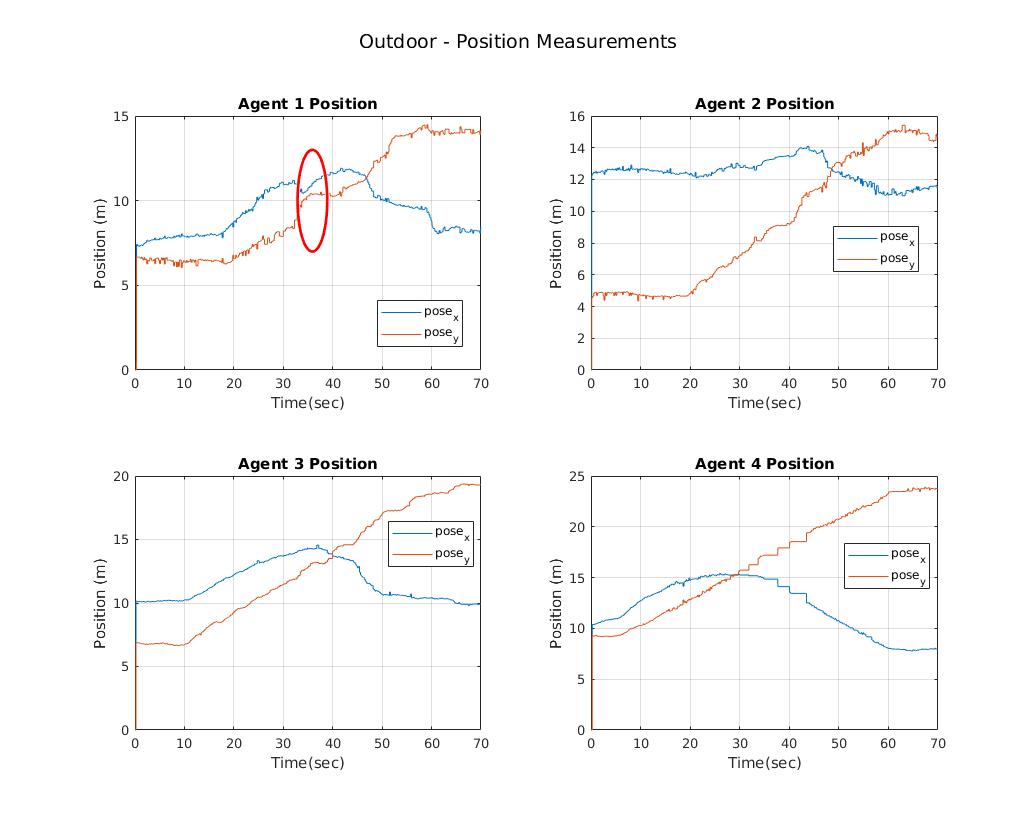}
    \caption{Pozyx position measurement data for agents from outdoor experiment with induced noise from windy outdoor conditions.}
    \label{fig:out_pose}
\end{figure}
 With wind speeds of approximately $10$ mph, Figures \ref{fig:out_vel_ref} and \ref{fig:out_pose} show the results of the outdoor experiment. There is some induced noise from the environmental disturbance evident in the position measurement data in Figure \ref{fig:out_pose}. However, the controller appears to behave appropriately, with agents 1 and 2 moving in and out of each other's collision radii (region in and around red circle in figure \ref{fig:out_pose} ). The initial and final state of the agents can be seen in figure \ref{fig:state}, where all agents have still maintained the original stable graph.

\section{Conclusions}

In this paper,
 we extended a framework we developed for studying stable motion and distributed topology control for multi-robot systems with directed interactions to the case of a multi-robot system with limited fields of view.  Then, we provided experimental results with a team of DJI Matrice 100 UAVs that demonstrated the effectiveness of the  control framework and showcased a  \textit{portable multi-robot experimental setup}. 

%
%
%

\bibliographystyle{IEEEtran}
\bibliography{ref}

\end{document}

%% file: macros.tex
\newcommand{\eqb}[1]{\begin{equation}\label{#1}}
\newcommand{\eqe}{\end{equation}}

\newcommand{\mbb}[1]{\mathbb{#1}}

\newcommand{\ib}{\begin{itemize}}
\newcommand{\ie}{\end{itemize}}

\newsavebox{\ieeealgbox}